\documentclass[iop,twocolumn]{emulateapj}

\newcounter{species}
\def\ion#1#2{\setcounter{species}{#2}#1$\;${\scriptsize\Roman{species}}\relax}

\newcommand{\teff}{T$_{\mathrm{eff}}$}
\newcommand{\na}{\ion{Na}{1}}
\newcommand{\ca}{\ion{Ca}{2}}
\newcommand{\ewna}{EW$_{\mathrm{NaI}}$ }
\newcommand{\ewca}{EW$_{\mathrm{CaII}}$ }

\usepackage{color}
\usepackage{natbib}
\usepackage{graphicx}
\usepackage{rotating}
\usepackage{epstopdf}
\usepackage{booktabs}
\usepackage{hyperref}

\shorttitle{A Near-Infrared Spectroscopic Survey of 886 Nearby M Dwarfs}
\shortauthors{Ryan C. Terrien et al.}
\begin{document}
\title{A Near-Infrared Spectroscopic Survey of 886 Nearby M Dwarfs}

\author{Ryan C. Terrien\altaffilmark{1,2,3,4},
Suvrath Mahadevan\altaffilmark{1,2,3}, 
Rohit Deshpande\altaffilmark{1,2}, 
Chad F. Bender\altaffilmark{1,2}
}
\slugcomment{ApJS Accepted, August 2015}

\email{rct151@psu.edu}
\altaffiltext{1}{Department of Astronomy and Astrophysics, The Pennsylvania State University, 525 Davey Laboratory, University Park, PA 16802, USA.}
\altaffiltext{2}{Center for Exoplanets and Habitable Worlds, The Pennsylvania State University, University Park, PA 16802, USA.}
\altaffiltext{3}{The Penn State Astrobiology Research Center, The Pennsylvania State University, University Park, PA 16802, USA.}
\altaffiltext{4}{Visiting Astronomer at the Infrared Telescope Facility, which is operated by the University of Hawaii under contract NNH14CK55B with the National Aeronautics and Space Administration.}

\begin{abstract}
We present a catalog of near-infrared (NIR) spectra and associated measurements for 886 nearby M dwarfs. The spectra were obtained with the NASA-Infrared Telescope Facility SpeX Spectrograph during a two-year observing campaign; they have high signal-to-noise ratios (SNR $>100-150$), span 0.8-2.4 $\mu$m and have $R\sim2000$. Our catalog of measured values contains useful \teff{} and composition-sensitive features, empirical stellar parameter measurements, and kinematic, photometric, and astrometric properties compiled from the literature. We focus on measures of M dwarf abundances ([Fe/H] and [M/H]), capitalizing on the precision of recently published empirical NIR spectroscopic calibrations. We explore systematic differences between different abundance calibrations, and to other similar M dwarf catalogs. We confirm that the M dwarf abundances we measure show the expected inverse dependence with kinematic, activity, and color-based age indicators. Finally, we provide updated [Fe/H] and [M/H] for 16 M dwarf planet hosts. This catalog represents the largest published compilation of NIR spectra and associated parameters for M dwarfs. It provides a rich and uniform resource for the nearby M dwarfs, and will be especially valuable for measuring Habitable Zone locations and comparative abundances of the M dwarf planet hosts that will be uncovered by upcoming exoplanet surveys.

\end{abstract}


\section{Introduction and Background}
Recent years have seen a crescendo of observational and theoretical interest in the population of nearby M dwarfs.  These stars, which make up the majority of the stars in the solar neighborhood \citep[e.g.][]{2002AJ....124.2721R}, have historically been difficult to observe due to their optical faintness. During the past two decades, great strides have been made in the refinement of techniques for identifying nearby M dwarfs. Multiple groups have used parallaxes \citep[e.g.][]{2006AJ....132.2360H,2011AJ....141...21W,Jao:2011er,Riedel:2014ce}, photometry, and proper motions to identify thousands of M dwarfs in the Solar neighborhood \citep{2005AJ....129.1483L,2011AJ....142..138L,Lepine:2013hc,2013MNRAS.435.2161F,Gaidos:2014if} and throughout the Galaxy \citep{2011AJ....141...97W,2013AJ....146..156D}.

In parallel, tools for observationally characterizing these stars have improved dramatically \citep{Reid:1995kw,Hawley:1996hg,Gizis:2002ej,2002AJ....124.2721R,2007ApJ...669.1235L}. Even so, the precise measurement of their stellar parameters remains challenging. Precise stellar radii and masses have only been measured for a small set of targets through interferometry \citep[e.g.][]{2009A&A...505..205D,Boyajian:2012eu} and the study of eclipsing binaries \citep[e.g.][]{2009ApJ...691.1400M,2011Sci...331..562C}. The measurement of M dwarf stellar compositions has also presented a formidable challenge, as the molecular species in their atmospheres mask the continuum \citep{1989ARA&A..27..701G}, typically rendering the direct spectroscopic modeling techniques intractable (although progress is being made, e.g.\ \citet{Onehag:2012hp}). Instead of direct modeling, the detailed analysis of atomic line and molecular absorption in M dwarf spectra has also been demonstrated \citep{2005MNRAS.356..963W,2006PASP..118..218W,2009PASP..121..117W}.

Rather than depending solely on the M dwarf spectra to derive abundance information, several groups have instead relied on abundance measurements of associated stars. These calibrator systems typically contain an FGK dwarf in a common-proper-motion system with an M dwarf, separated by several arcseconds or more. The composition of the higher-mass star is known or readily determined using established spectroscopic techniques \citep[as deployed for M subdwarfs in][]{1997AJ....113..806G}. The M dwarf is then assumed to have the same composition, and their photometric and spectroscopic properties are used to construct empirical models that can then be applied to single stars. This empirical calibration strategy was first implemented for optical photometry \citep[$V-K_s$,][]{2005A&A...442..635B}, taking advantage of the strong and weak [Fe/H] dependence for M dwarf $V$ and $K_s$-band absorption, respectively \citep{2000ApJ...542..464C,Delfosse:2000ur,2005A&A...442..635B}. Multiple groups have developed similar photometric calibrations based on absolute stellar magnitudes \citep{2009ApJ...699..933J,2010A&A...519A.105S,2012A&A...538A..25N,2012AJ....143..111J} and colors \citep{2014AJ....147...20N,Hejazi:2015fs}, and \citet{Neves:2013cm,Neves:2014jj} have further leveraged these photometric [Fe/H] calibrations into more precise optical spectroscopic calibrations.

This empirical calibration technique has also been implemented for near-infrared (NIR) spectra, taking advantage of the relative brightness of M dwarfs in the NIR. The first of these was the $K_s$-band calibration developed by \citet{RojasAyala:2010ht}, which leveraged the $K_s$-band \ion{Na}{1} doublet and \ion{Ca}{1} triplet, along with H$_2$O absorption indices to account for \teff{} effects. This technique was expanded by \citet[][, T12]{2012ApJ...747L..38T}, refined in \citet{RojasAyala:2012fb} and \citet{2014AJ....147...20N} (hereafter N14), and further generalized in \citet{Mann:2013kj} (hereafter M13a). Presently, the NIR calibration with the largest number and diversity of calibrators is that of M13a for early-mid M dwarfs, and \citet{Mann:2014bz} (hereafter M14) for late-type M dwarfs; these calibrations provide [Fe/H] and [M/H], are well-calibrated across a wide range of [Fe/H] ($-.8 <$ [Fe/H] $< .5$) and spectral type (late K to M8), and have good precision ($\sim0.10$ dex in [Fe/H] and [M/H]). 

In addition to the empirical techniques for measuring stellar composition, various groups have  developed similar techniques for the measurement of spectral type, effective temperature, stellar mass, and stellar radius, based on spectral type standards, stars with interferometrically-determined parameters, or stars with parameters measured through other means. Notably, techniques applicable to high resolution optical spectra from HARPS \citep{Neves:2013cm,Neves:2014jj,Maldonado:2015wk} and Keck \citep{Pineda:2013wy} have achieved good precision for the M dwarfs observed in these programs. Techniques based on lower-resolution ($R\sim2000$) NIR spectra, using \teff{}-sensitive indices \citep{Mann:2013fv} (hereafter M13b) and atomic absorption features \citep{Newton:2015ir} have been also been developed. The use of broadband photometry has also enabled precise measurements of the temperatures and radii of these stars \citep{2015ApJ...804...64M}. The resulting suite of techniques has been used extensively in the characterization of \textit{Kepler} low-mass planet hosts \citep{Muirhead:2012dxa,Muirhead:2012ja,2012AJ....143..111J,Ballard:2013ir,Muirhead:2013ft,Muirhead:2014gw,Newton:2015ir}.

The tools and data described above, combined with advances in structural \citep{Baraffe:1998ux,Dotter:2007ht,Spada:2013cs} and atmospheric \citep{1997ARA&A..35..137A} models, provide a fertile ground for a variety of topics related to M dwarfs, including Galactic evolution \citep[e.g.][]{2010AJ....139.2679B,2011AJ....141...98B}, stellar structure \citep[e.g.][]{LopezMorales:2007ea,Reiners:2012ew,2012ApJ...757...42F,Torres:2013ib}, and exoplanets. Nearby M dwarfs are appealing targets for Doppler Radial Velocity (RV) and transit exoplanet searches because their low masses and luminosities make planets orbiting in their Habitable Zones \citep[HZ][]{Kasting:1993hw,2013ApJ...765..131K,Kopparapu:2014fs} more easily detectable than those around higher-mass stars. Moreover, given the prevalence of M dwarfs, their planets may represent the most common exoplanetary environment in the Galaxy. Upcoming NIR ground-based RV surveys such as the Habitable-zone Planet Finder on the Hobby-Eberly Telescope \citep[HPF,][]{2012SPIE.8446E..1SM}, CARMENES \citep{Quirrenbach:2012ir} at Calar Alto, SPIrou on the CFHT \citep{Thibault:2012fb}, IRD on Subaru \citep{Kotani:2014fe}, and iLocater on LBT as well as photometric surveys like TESS \citep{Ricker:2015ie}, PLATO \citep{Rauer:2014kx}, and MEarth \citep{2008PASP..120..317N} will specifically target planets around nearby M dwarfs. These surveys will provide unprecedented sensitivity to planets orbiting the nearby M dwarfs, whose optical faintness has made them difficult targets for many existing RV and transit surveys. Despite being optimized for optically bright FGK stars, results from \textit{Kepler} \citep{2010Sci...327..977B} clearly indicate that M dwarfs host a large population of small planets in the HZ \citep{2015ApJ...807...45D}. The characteristics of planets around M dwarfs will be useful for testing theories of planet formation \citep[e.g.][]{Laughlin:2004hf} in the regime of low stellar and protoplanetary disk masses.

In order to construct a well-characterized target list for HPF, we have carried out  an extensive spectroscopic survey of 886 nearby M dwarfs, using the SpeX spectrograph at the NASA Infrared Telescope Facility (IRTF). We take advantage of the NIR spectroscopic techniques described above to measure the metallicities of these stars, in order to form the largest catalog yet of nearby M dwarf compositions based on spectroscopy. We also provide measurements of the \teff{}-sensitive indices, which can be used to estimate stellar mass, radius, or luminosity. This work echoes the spirit of previous large-scale efforts to characterize our most common stellar neighbors \citep[e.g.][N14]{Reid:1995kw,Riaz:2006du,2013AJ....146..156D}.

In Section \ref{obs} we describe our observations and data reduction, including target selection and cross-identification with other catalogs. In Section \ref{specmeas} we describe our spectral measurements and error estimates. In Section \ref{gendisc} we provide a brief summary and discussion of our measurements and comparisons to other catalogs.

\subsection{Published Results from this Sample}
\label{secpub}
This dataset has already enabled a variety of studies of the nearby M dwarfs. T12 developed a new $H$-band [Fe/H] calibration for M dwarfs, extending established techniques \citep{2005A&A...442..635B,2006PASP..118..218W,RojasAyala:2010ht} and providing a valuable comparison for the development of improved calibrations (M13a) and a starting point for work with higher resolution spectra \citep[e.g. SDSS-III APOGEE or IGRINS,][]{2013AJ....146..156D,Park:2014kn}. \citet{Terrien:2012hj} employed the T12 [Fe/H] measurement technique to characterize the low-mass eclipsing binary CM Draconis \citep[CM Dra][]{2009ApJ...691.1400M}, helping to remove the [Fe/H]-related degeneracy and to clarify the disagreement between models and observations of this system. \citet{Terrien:2014jq} again applied the T12 [Fe/H] measurement technique to confirm the first low-mass members of the nearby Coma Berenices cluster \citep{Trumpler:1938vy}. Finally, \citet{Terrien:2015eu}  leveraged M dwarf activity catalogs \citep[e.g.][]{Gizis:2002ej,Riaz:2006du,Gaidos:2014if} and improvements in parallax measurements of the nearby M dwarfs \citep[e.g.][and the Research Consortium on Nearby Stars, RECONS\footnote{\url{http://www.recons.org}}]{Dittmann:2014cr} to develop new spectroscopic indices for M dwarf luminosity, surface gravity, and possibly $\alpha$-enrichment. 

\section{Observations and Data Reduction}
\label{obs}
Over a period of two years, beginning in 2011 May and concluding in 2013 August, we observed 886 M dwarfs with the SpeX NIR spectrograph on the NASA-IRTF \citep{2003PASP..115..362R}. This set of observations includes 21 stars observed multiple times. 

\subsection{Target Selection}
We drew M dwarf targets primarily from the \citet[][LSPM-N]{2005AJ....129.1483L} and \citet[][LG11]{2011AJ....142..138L} proper-motion based catalogs of nearby M dwarfs, and from published lists of M dwarf planet hosts\footnote{e.g.~\url{http://exoplanets.org}} and known wide binaries \citep[drawn from ][]{RojasAyala:2010ht} that serve as abundance calibration stars. The majority of target stars (829) were taken from LG11, the remainder being the aforementioned binary calibrator stars, planet hosts, or that were targeted in a contemporaneous high-resolution survey \citep{2013AJ....146..156D}. For stars from LG11 and LSPM-N, we prioritized targets with redder optical-to-infrared color ($V-J > 4$), a good indication of mid-late M spectral subtype (M4 or later). The resulting compilation of targets included both high-priority targets (calibrator stars, planet hosts, late-type M dwarfs) and low-priority targets (early type M dwarfs).

From this super-set of potential targets we constructed our nightly observing plans, and during each night we actively added and removed targets to account for changes in observing conditions. Our primary goal was to ensure very high signal-to-noise (SNR $> 200$ per pixel) for high-priority targets, and our secondary goal was to observe as many M dwarfs as possible with reasonable SNR ($>100$). The typical resulting observing pattern was a series of groups (in position and time) which contained a few faint late-type (M4 or later) M dwarfs and several brighter early (earlier than M4) M dwarfs. Each group was sufficiently localized that a single nearby standard star (within one hour of time and 0.1 $\sec{z}$ of target star) could be used for telluric correction, and the same arc lamp exposure could be used for wavelength calibration. This strategy enabled a large number of M dwarfs to be observed, with minimal time losses from large slews, wavelength calibrations, and standard star observations. 

\subsection{Data Reduction}
\label{datared}
We obtained uniform data for all observations, operating SpeX in the short cross-dispersed (SXD) mode with the $0.3 \times 5''$ slit, which produces $R \sim 2000$ spectra from 0.8 - 2.4 $\mu\rm{m}$. As described in T12, we extracted these spectra with the facility-provided SpeXTool package \citep{2004PASP..116..362C}. We telluric-corrected using the \texttt{xtellcor} program \citep{Vacca:2003fw} and observations of an A0V star (or similar), at an airmass and observation time within 0.1 ($\sec{z}$) and 1 hour of the target, respectively. The \texttt{xtellcor} program also flux-calibrates the spectrum, using the  optical magnitudes (from SIMBAD) of the standard star to estimate its flux as a function of wavelength.  \citet{Vacca:2003fw} estimate that the flux calibration using this method agrees to within a few percent of fluxes derived using broadband magnitudes.

\begin{figure*}
\begin{center}
\includegraphics[scale=0.5]{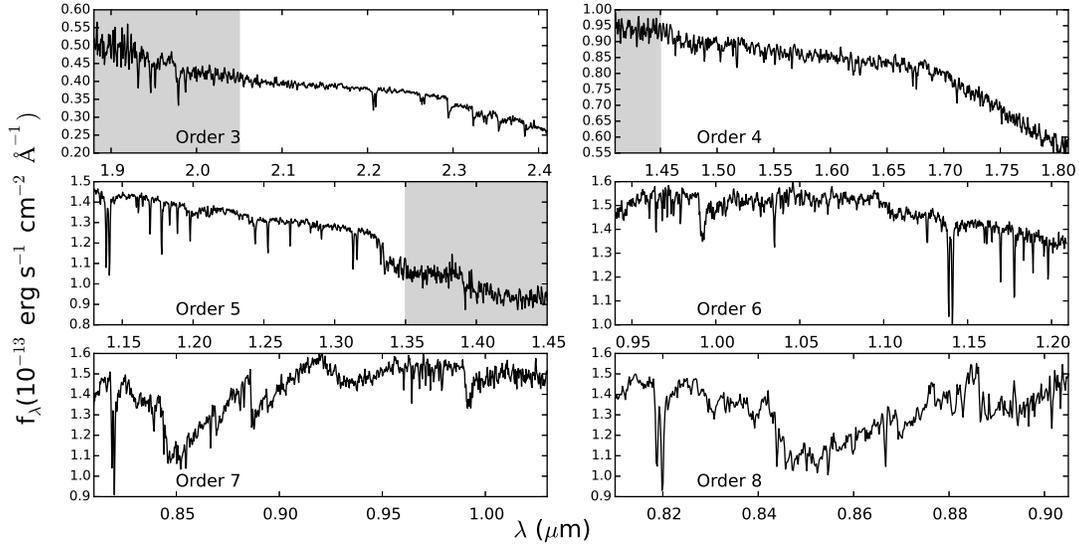}
\caption{The six separate telluric-corrected orders of the IRTF-SpeX SXD spectrum of GJ 1103A, which is a typical M4.5V star from our survey. Regions with significant telluric contamination are shaded in gray.
\label{specex}}
\end{center}
\end{figure*}

A typical example spectrum is shown in Figure \ref{specex}. All observations presented here took place prior to the SpeX detector upgrade, which occurred in the spring of 2014.

\subsection{Cross-Identification with other catalogs}
We cross-identified our targets with several published catalogs, in order to provide supplementary data for our target stars and to facilitate the use of our catalog,  Each target has a 2MASS \citep{Skrutskie:2006hl} identifier, and we use the 2MASS J2000 coordinates in our final catalog. The 2MASS ID also provides the most convenient uniform identifier for all our targets. We include in our catalog the 2MASS $J,H,K_{s}$ photometry and the respective quality flags for each star.

We cross-identified our targets with the UCAC4 \citep{Zacharias:2013cf} catalog as well. UCAC4 provides optical magnitudes based on the AAVSO Photometric All-Sky Survey (APASS). Optical magnitudes are useful for M dwarfs as their optical-to-infrared colors are excellent indicators of spectral type and \teff{}; we include UCAC4 aperture ($A$)-magnitudes (for 874 stars) and $V$-magnitudes (for 732 stars) as indications of their optical brightness. 

For the majority of our targets, we include proper motions as measured by LG11 or LSPM-N. For targets without proper motions in these catalogs, we include measurements from UCAC4 or the PPMXL catalog \citep{Roeser:2010cr}.

We also provide parallax measurements for a nearly half of our targets (427). Many of these parallax measurements are drawn from the RECONS database \citep{Jao:2005dj,Riedel:2014ce,Riedel:2010dy,2006AJ....132.2360H} and the newly-published MEarth parallax catalog \citep{Dittmann:2014cr}. We also found parallaxes in several other catalogs \citep{vanAltena:1995ts,vanLeeuwen:2007dc,Gould:2004jo,AngladaEscude:2012gk,Harrington:1993jc,Gatewood:2008hu,Monet:1992bs,2009AJ....137.4109L}.

\section{Spectral Measurements}
\label{specmeas}
We describe here the various spectral measurements we performed on the M dwarf spectra.

\subsection{Band Indices and Spectral Type}
\label{section:bands}
The NIR spectra of M dwarfs are dominated by the broad absorption of H$_2$O, the strength of which is known to correlate strongly with \teff{} and spectral type \citep[e.g.][]{Wilking:1999ge}. Using $R\sim2000$ NIR spectra like ours, several groups have developed H$_{2}$O-based and other indices that are well-correlated with spectral type and \teff{} \citep[][T12,M13a, M13b, N14, M14]{2010ApJ...722..971C,RojasAyala:2010ht,RojasAyala:2012fb}. Calibrations of these indices provide an efficient and simple method for measuring spectral type and temperature for the M dwarfs in our catalog. The relatively large widths of the wavelength ranges used in these measurements makes them insensitive to small errors in the RV measurement. For the M dwarfs in our catalog, we measured the \teff{}-sensitive indices used in M13b and N14 \citep{2010ApJ...722..971C,RojasAyala:2012fb,Mann:2013kj} (see Table \ref{h2oparams}).

\begin{deluxetable*}{lccc}
	\tablewidth{0pc}
	\tablecaption{Parameters of \teff{}-Sensitive Indices\tablenotemark{a} \label{h2oparams}}
	\tablehead{\colhead{Name} &  \colhead{Range 1 (\AA)}  & \colhead{Range 2 (\AA)} & \colhead{Range 3 (\AA)}} \\
	\startdata
	\citet{Mann:2013fv} $J$ Index & $ 10087.5 - 10172.5 $ & $ 10967.5 - 11052.5 $ & $ 11587.5 - 11672.5 $  \\
    \citet{Mann:2013fv} $H$ Index & $ 14625.0 - 14655.0 $ & $ 16615.0 - 16645.0 $ & $ 17945.0 - 17975.0 $  \\
    \citet{Mann:2013fv} $K$ Index& $ 20242.5 - 20277.5 $ & $ 22747.5 - 22782.5 $ & $ 23367.5 - 23402.5 $  \\
	\citet{Mann:2014bz} H$_2$O-$J$ & $ 12100.0 - 12300.0 $ & $ 13130.0 - 13330.0 $ & $ 13310.0 - 13510.0 $  \\
    \citet{2010ApJ...722..971C} H$_2$O-$H$ & $ 15950.0 - 16150.0 $ & $ 16800.0 - 17000.0 $ & $ 17600.0 - 17800.0 $  \\
    \citet{RojasAyala:2012fb} H$_2$O-$K$ & $ 20700.0 - 20900.0 $ & $ 22350.0 - 22550.0 $ & $ 23600.0 - 23800.0 $  \enddata
    \tablenotetext{a}{Value of index $= (\langle F_1 \rangle - \langle F_2 \rangle) / (\langle F_2 \rangle - \langle F_3 \rangle)$}
\end{deluxetable*}

To measure spectral type for the M dwarfs in our catalog, we applied the NIR calibration developed in N14, which is based on the H$_{2}$O-K2 index developed in \citet{RojasAyala:2012fb}. This calibration yields ``NIR M sub-types" on a uniform scale for our stars. The precise relationship between this NIR M sub-type and the established Palomar/Michigan State University (PMSU) scale \citep{Reid:1995kw,Hawley:1996hg} is considered in \citet{RojasAyala:2012fb} and N14, and is known to be sensitive to [Fe/H] for at least the early M dwarfs. We also include in our catalog the measured PMSU spectral types from the N14 PMSU spectral type calibration over the range in which it is valid (M1-M4).

To check the consistency of our NIR spectral type measurements, we considered the overlap between our sample and that of N14, which includes 152 M dwarfs. We found no significant offset between the respective measures of spectral type, and a scatter of half a subtype, consistent with the scatter expected from N14.

This NIR spectral type measurement was necessarily the first in our chain of measurements, as it defined the template for our measurement of RV. The \teff{}-sensitive features are sufficiently broad that they are insensitive to RV shifts in our $R\sim2000$ spectra. Nonetheless, after measuring the RV we re-measured the spectral type in the rest frame of the star, and this is the value we report in our catalog. 

\subsection{Radial Velocity}
We measured the RV of each order of each spectrum by fitting a Gaussian to the cross-correlation (in log-$\lambda$) of the observed spectrum with a spectral template of similar spectral type. We used three spectral templates from the IRTF Cool Stars Library \citep{2005ApJ...623.1115C,2009ApJS..185..289R}: for stars M4 and earlier we used the M1.5V star HD 36395, for stars M5 to M6 we used the M5V star Gl 51, and for stars later than M6 we used the M9V star LHS 2065. For each observation, we report the weighted (by SNR) average RV from all orders, and this is value we used to shift all spectra to a common frame for further analysis.

\begin{figure}
\begin{center}
\includegraphics[scale=0.35]{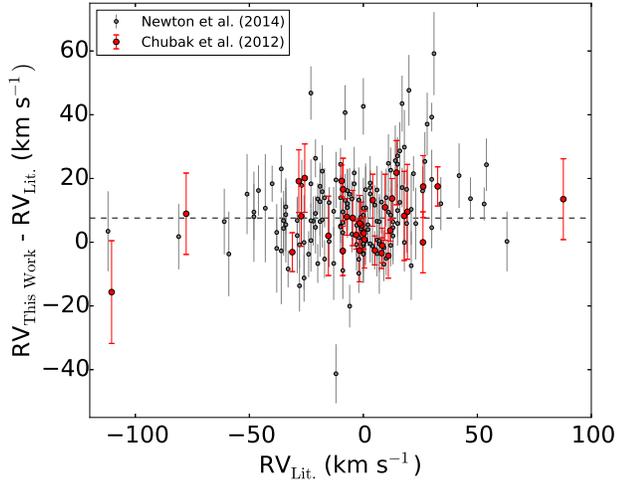}
\caption{Differences between the RVs measured in this work and those measured in \citet{Chubak:2012tv} and N14. The median differences are shown as dashed lines, and are both approximately 8~km~s$^{-1}$. 
\label{rv_validation}}
\end{center}
\end{figure}

To verify our RV measurements, we compared our (barycentric-corrected) RV measurements to those measured in \citet{Chubak:2012tv} and N14. For the 33 stars in common with \citep{Chubak:2012tv}, we find a median offset of $+7.6$~km~s$^{-1}$ with a standard deviation of $8.7$~km~s$^{-1}$. For 152 stars in common with N14, we find a median offset of $+7.9$~km~s$^{-1}$ with a standard deviation of 13.2~km~s$^{-1}$. Figure \ref{rv_validation} shows the differences in RV between this work and the literature sources. Among our common targets with the N14 study, there were seven with differences greater than 40~km~s$^{-1}$. Two of these (2MASS J07362513+0704431/GJ 3454 and 2MASS J17074083+0722066/GJ 1210) are known visual binaries with sub-arcsecond separation \citep{Pravdo:2005kt,Janson:2014fx}. Given the consistency of the RV measurements of the majority of the common targets, we suspect that these highly discrepant targets are either RV-variable or binary stars. We also measure a large RV for one star (2MASS J06052936+6049231/LHS 1817, $-113.9$~km~s$^{-1}$) with similar magnitude but the opposite sign as N14 and \citet{Shkolnik:2012cs}. 

That the systematic offset from N14 and \citet{Chubak:2012tv} would be similar is unsurprising, as N14 implement a $\sim2.6$~km~s$^{-1}$ correction to match the RV system of \citet{Chubak:2012tv}. The origin of the $\sim7-8$~km~s$^{-1}$ offset for our data is unclear; since we rely solely on the arc lamp wavelength calibrations, we are subject to different and possibly worse instrumental systematics than N14, who use telluric spectra to refine their wavelength calibrations. A systematic issue with the wavelength calibration of approximately 8~km~s$^{-1}$ is also supported by the magnitudes of the inter-order and RV template differences, as discussed in Section \ref{uncertainties}.

In order to obtain barycentric RVs in the same system as \citep{Chubak:2012tv} and N14, the median offset above ($\sim$7.6~km~s$^{-1}$) should be added to the RVs in our catalog.

\subsection{Effective Temperature and Related Parameters}
Due to the difficulty of modeling M dwarf atmospheres and spectra, model-dependent measurements of \teff{} can systematically differ by 100 K or more \citep[][and references therein]{Rajpurohit:2013ij}. Interferometric \teff{} measurements have therefore been crucial in defining the M dwarf \teff{} scale, and in parsing the relations between the fundamental parameters of M dwarfs \citep[][and references therein]{Boyajian:2012eu}. However, there are few targets with spectral type M5 or later that are bright enough or have large enough angular diameters to support precise interferometric radius measurements. To enable stellar radius measurements without interferometry, multiple groups have constructed empirical calibrations based on photometry \citep{Boyajian:2012eu}, spectral indices (M13b), or individual absorption feature strengths \citep{Newton:2015ir}. 

We measured \teff{} for the M dwarfs in our catalog using the H$_{2}$O indices as discussed in Section \ref{section:bands}. We used the  $J,H,K_s$ \teff{} calibrations developed in M13b over their well-calibrated range (e.g.\ $1.0 < \mathrm{H}_2\mathrm{O-}K < 1.37$), which corresponds to $3300 < \mathrm{T}_{\mathrm{eff}} < 4800$~K. M13b provide empirical relations for stellar radius, luminosity, and mass as a function of \teff{}, calibrated on a set of stars with interferometrically measured parameters. We favor the use of the $K_s$-band index and calibrations due to the robustness of this index to small changes in the telluric correction (as discussed in Section \ref{section:bands}). 

We also measured \teff{}, stellar radius, and stellar luminosity using the $H$-band atomic feature strengths studied in \citet{Newton:2015ir}. We measure the pseudo-equivalent widths (EWs) of these features using the continuum and feature definitions in their Table 1. They construct relations using the strengths and ratios of Mg and Al lines in M dwarfs, empirically calibrated on stars with interferometrically measured parameters. These relations use a similar calibration set to that of M13b, so the range of validity is similar: $3200 < T_{\mathrm{eff}} < 4800$~K.

\subsection{Composition}
\label{secmeas:comp}
We measured [Fe/H] and [M/H] using the empirical spectroscopic calibrations from T12, M13a, M14, and N14. These calibrations were all developed using data from the same spectrograph (IRTF-SpeX) in the same settings (SXD) as ours, so their recipes can be applied directly to our data. M13a present separate calibrations for features in the $J,H,K_s$ bands that yield both [Fe/H] and [M/H] for early-mid M dwarfs (M1-M5). They derived these calibrations from a blind search of the most [Fe/H] and [M/H]-sensitive regions of the spectra in 112 stars with known [Fe/H] and [M/H], and are based on the EWs of a small number (four or fewer) of spectral regions in each case. The calibrations are valid for K5-M5 dwarfs and over approximately $-1.0 < $[Fe/H] $< +0.56$. We report the results of these calibrations separately for each target in our catalog. 

For a similar range of targets in spectral type and [Fe/H], we also report the $K_s$-band [Fe/H] calibration developed in N14, which depends only on the EW of the $K_s$-band \ion{Na}{1} doublet. N14 calibrated this relation using 36 M dwarfs with known [Fe/H], and it is similar in form to the relations developed in \citep{RojasAyala:2010ht} and \citep{RojasAyala:2012fb}. It is valid for M1-M5 dwarfs and for a range of $-1.00 < $[Fe/H]$ < +0.35$.

For the late M dwarfs (M5 or later), the only applicable published [Fe/H] calibration is that of (M14). The M14 calibration is based on two strong features in the $K_s$-band, the \ion{Ca}{1} triplet and the \ion{Na}{1} doublet, which were also used in the calibration of \citet{RojasAyala:2010ht} and \citet{RojasAyala:2012fb} for early-type M dwarfs. This calibration is based on 44 M dwarfs with known metallicities. It is an extension of the binary calibration technique described above, which extends the calibration set by including some early+late M dwarf pairs where the earlier-type M dwarf [Fe/H] has itself been measured using empirical calibrations. This late-type calibration is valid for M4.5-M9.5 dwarfs and for a range of $-0.58 <$[Fe/H]$ < +0.56$. The summary of the parameters of this [Fe/H] calibration and those above is shown in Table \ref{calparams}. 

For general comparison with other catalogs, and for cases below in which it is useful to have a single [Fe/H] measurement for a given star, we elect to use the $K_s$-band calibrations of M13a for M1-M5 (see Section \ref{prefmeas}), along with the output of M14 for M5 and later. We also report the [M/H] values where available. For the purposes of presenting and vetting our catalog, we primarily consider [Fe/H] as this is the specific quantity addressed by the bulk of work on the topic of M dwarf compositions.

\begin{deluxetable*}{lccc}
\tablewidth{0pc}
\tablecaption{Parameters of primary empirical calibrations \label{calparams}}
\tablehead{\colhead{Calibration} &  \colhead{SpT Range}  & \colhead{Metallicity Range} & \colhead{Error} } \\
\startdata
\citet{Mann:2013kj} [Fe/H]$_{K}$ & K5.0$-$M5.0  & $-1.04 <\mathrm{[Fe/H]}<+0.56$ & 0.11 dex \\
\citet{Mann:2013kj} [M/H]$_{K}$ & K5.0$-$M5.0  & $-0.70 <\mathrm{[M/H]}<+0.50$ & 0.10 dex \\
\citet{Mann:2014bz} [Fe/H]$_{K}$ & M4.5$-$M9.5  & $-0.58 <\mathrm{[Fe/H]}<+0.56$ & 0.07 dex \\
\citet{2014AJ....147...20N} [Fe/H] & M1.0$-$M5.0  & $-1.00 <\mathrm{[Fe/H]}<+0.35$ & 0.12 dex  \\
\citet{2014AJ....147...20N} NIR SpT & M1.0$-$M9.0  & \ldots & 0.5 subtype \enddata
\end{deluxetable*}

\subsection{Activity and gravity-sensitive lines}
\label{agfeats}
Our spectra extend to approximately 0.8 $\mu$m on the blue end, and so contain the 8200\AA\,\ion{Na}{1} doublet (hereafter \na{}) and the 8600\AA\,\ion{Ca}{2} triplet (hereafter \ca{}), strong features that have shown promise as indicators of activity \citep[e.g.][]{Kafka:2006kw,Barnes:2014bj} and surface gravity \citep{Schlieder:2012iw}. Our measurements of these features for a subset of the observed M dwarfs showed a strong relationship with the \teff{} and [Fe/H] measures in these stars, and can be used to derive information about their radii, luminosities, and possibly $\alpha$-enrichment. These measurements are discussed in \citep{Terrien:2015eu}; we repeat the salient details here. 

We measured the EWs of the \na{}~and \ca{}~features (\ewna{}, \ewca{}) for each of our targets. For \na{}, we chose line and pseudo-continuum regions similar to those used in \citet{Martin:2010cx}, \citet{Schlieder:2012iw}, and \citet{GalvezOrtiz:2014jb}. For \ca{}, we defined feature regions for each of the three lines, as well as pseudo-continuum regions blueward and redward of each line. In each case, we fit a line to the pseudo-continuum regions in order to define the pseudo-continuum, and calculated the feature EW relative to this line. For \ewna{} and \ewca{} we follow the convention that positive EW corresponds to line absorption. 

\begin{figure*}
\begin{center}
\includegraphics[scale=0.5]{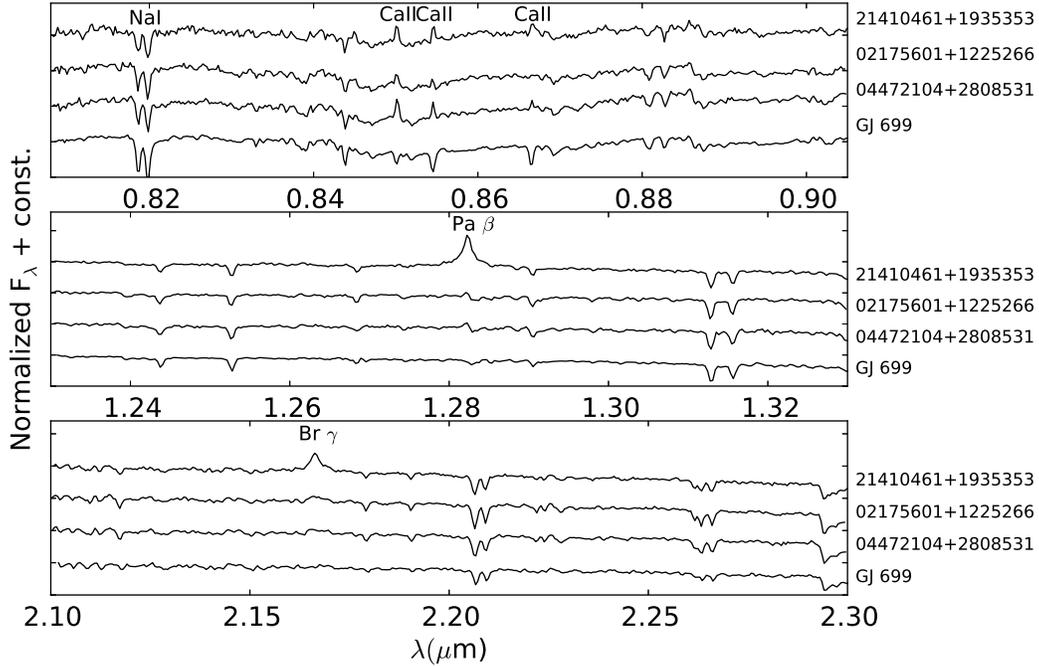}
\caption{Three stars in the survey sample that have strong emission lines, along with GJ 699 as an example of a typical M dwarf. Three spectral regions with strong emission lines are shown: the top highlights the NIR Ca triplet, the middle highlights the H Pa $\beta$ line, and the bottom highlights the H Br $\gamma$ line. The top panel also identifies the \na{} and \ca{} features described in Section~\ref{agfeats}.
\label{emission}}
\end{center}
\end{figure*}

We caution that the \ca{} and \na{} features are embedded in regions of strong TiO and other absorption (as can be seen in e.g.\ \citet{2009ApJS..185..289R}), so the EWs we measure are likely strongly sensitive to effects of the line species themselves as well as to the species that define the local pseudo-continuum. Moreover the SNR in the relevant orders of the SpeX spectra (orders seven and eight) is typically only a third of that in the $J,H,K_s$ bands. Accordingly, these feature strengths should be carefully vetted before they are used in any application. The regions we defined for this analysis are listed in Table \ref{canaparams} and are shown on example spectra in Figure \ref{emission}. Figure \ref{emission} also shows the three (possibly young) M dwarfs in our sample for which the \ca{} lines are in emission.

\begin{deluxetable}{lcccc}
\tablewidth{0pc}
\tablecaption{Definitions for 8200\AA\,\ion{Na}{1} and 8600\AA\,\ion{Ca}{2} \label{canaparams}}
\tablehead{\colhead{Feature} & \colhead{Center} & \colhead{Width} &  \colhead{PC Blue\tablenotemark{a}} & \colhead{PC Red\tablenotemark{a}} \\
 & \colhead{(\AA)} & \colhead{(\AA)} & \colhead{(\AA)} & \colhead{(\AA)} 
}\\
\startdata
\ion{Na}{1} & 8192 & 34.0 & 8149 - 8169 & 8230 - 8270 \\
\ion{Ca}{2}$_1$ & 8500 & 12 & 8456 - 8488 & 8512 - 8536 \\
\ion{Ca}{2}$_2$ & 8544 & 12 & 8512 - 8536 & 8553 - 8573 \\
\ion{Ca}{2}$_3$ & 8664 & 12 & 8635 - 8652 & 8672 - 8688 \enddata
\tablenotetext{a}{PC = Pseudo-continuum region definition.}
\end{deluxetable}

\subsection{Estimation of Uncertainties}
\label{uncertainties}
\begin{figure*}
	\begin{center}
		\includegraphics[width=.8\paperwidth]{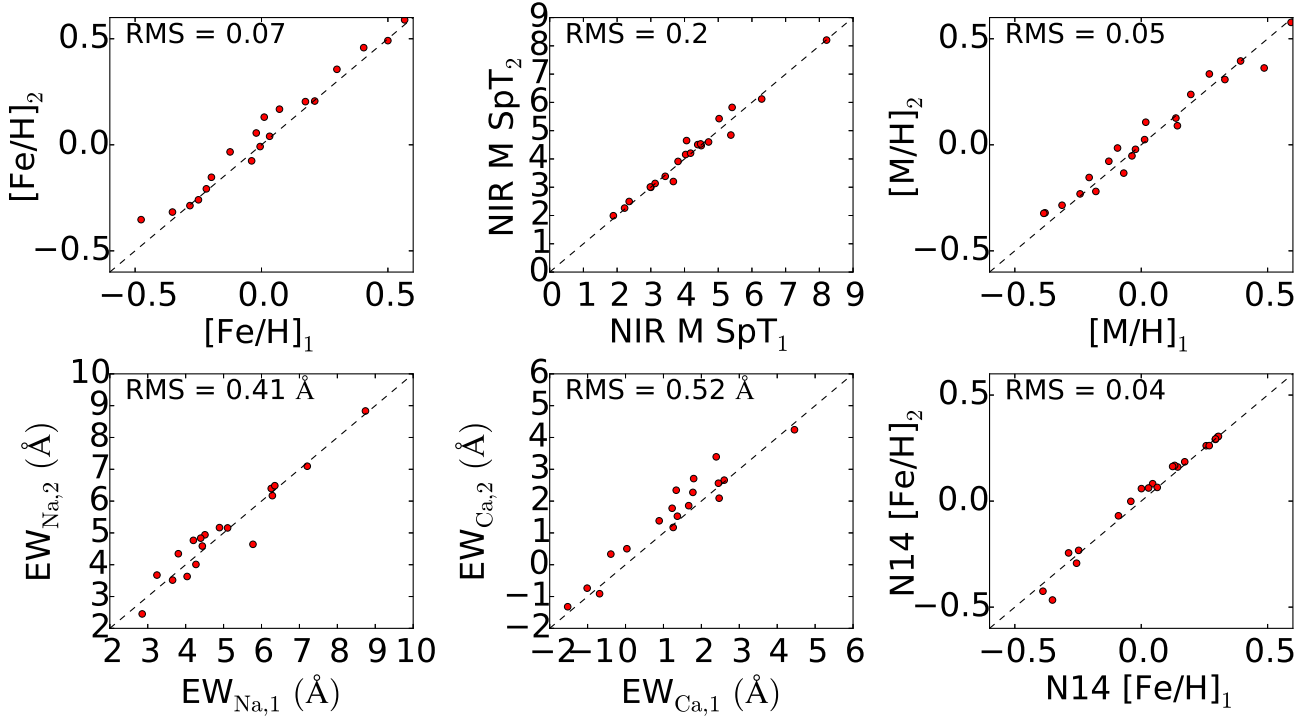}
		\caption{For targets we observed multiple times, the comparison of the [Fe/H] (including our primary [Fe/H] as discussed in Section \ref{secmeas:comp} and using the technique of N14), [M/H], M subtype, and \ca{} and \na{} feature strength measurements, from two different epochs. The black lines indicate a one-to-one relation. We adopt the observed RMS scatter in each case as our measurement uncertainty.
			\label{errcomp}}
	\end{center}
\end{figure*}

\begin{figure*}
	\begin{center}
		\includegraphics[width=.8\paperwidth]{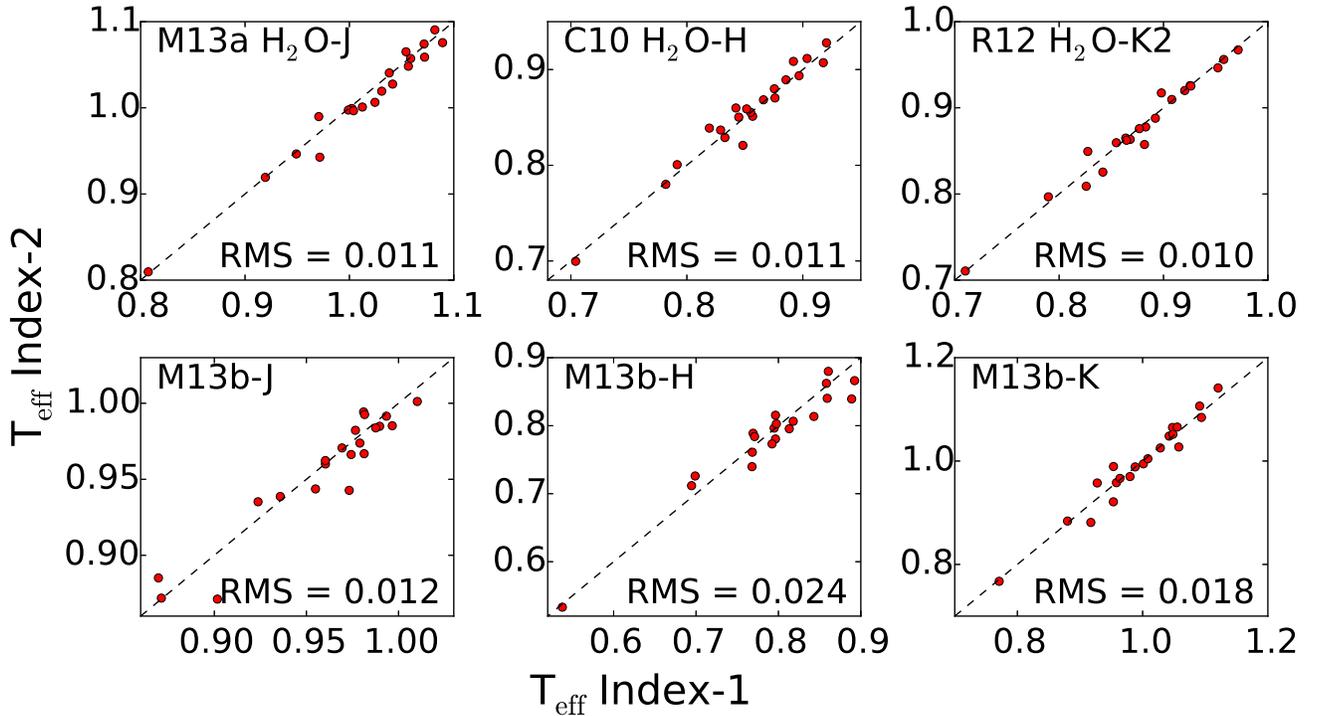}
		\caption{For targets we observed multiple times, the comparison of the \teff{}-sensitive index measurements from two different epochs. N14 refers to the bands defined in N14, and M13 refers to those defined in M13b. The black lines indicate a one-to-one relation. We adopt the observed RMS scatter in each case as our measurement uncertainty.
			\label{h2oerrcomp}}
	\end{center}
\end{figure*}

\begin{figure*}
	\begin{center}
		\includegraphics[width=.8\paperwidth]{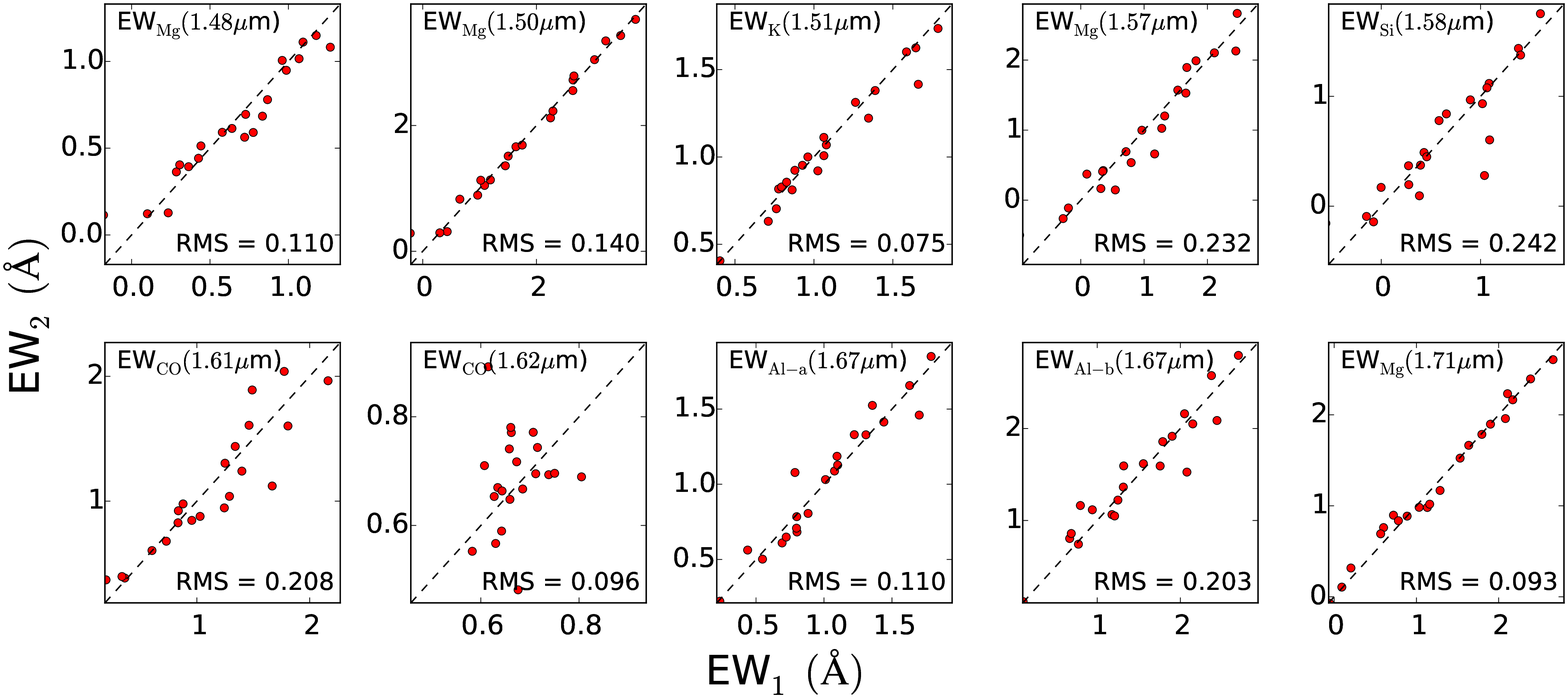}
		\caption{For targets we observed multiple times, the comparison of the \citet{Newton:2015ir} feature strength measurements from two different epochs. The black lines indicate a one-to-one relation. We adopt the observed RMS scatter in each case as our measurement uncertainty.
			\label{nferrcomp}}
	\end{center}
\end{figure*}

In order to estimate our measurement uncertainties (for all quantities other than RV), we rely on a set of 21 M dwarfs with multiple observations. We calculate the RMS of the epoch-to-epoch differences in each measurement to obtain an observational estimate of our internal uncertainty. These values are shown for [Fe/H], [M/H], M spectral sub-type, \na{}, and \ca{} in Figure \ref{errcomp}, for the \teff{}-sensitive indices in \ref{h2oerrcomp}, and for the $H$-band features of \citet{Newton:2015ir} in Figure \ref{nferrcomp}. For the \ca{} and \na{} features, we exclude three pairs with very low SNR ($<60$). The observed RMS differences naturally include effects that arise from variable telluric conditions and by-hand choices in the telluric correction pipeline. The resultant variable quality of the telluric correction can induce correlated noise which dominates our errors at the high SNRs we achieve (N14).

We note that the \teff{}-sensitive indices may be affected by the broad characteristics of the telluric correction, including the flux calibration (which uses SIMBAD $B,V$ magnitudes, Section \ref{datared}) and the construction of the telluric spectrum. By repeating the telluric correction with varying $B,V$ magnitudes for a small set of stars, we found that the $J$-band indices were generally much more sensitive to differences in $B,V$ magnitude and small differences in the by-hand components of the telluric correction. We favor the use of the $K_s$-band indices (H$_2$O-K2 or the M13b $K_s$-index), which are stable to a few percent or better with $B,V$ variations of 0.5~mag.

The \teff{}-sensitive indices can be used to estimate stellar \teff{} (also mass and bolometric luminosity) as described in M13b, who present uncertainties of $70-100$~K for these relations. Our adopted measurement uncertainties for their \teff{}-sensitive indices ($\sim0.01-0.02$) correspond to median uncertainties of 110~K ($J$), 170~K ($H$), and 78~K ($K_s$) when propagated through the \teff{}-index relations from M13b.

The $H$-band atomic feature strengths presented in \citet{Newton:2015ir} can similarly be used in empirical calibrations to derive stellar \teff{} (residual scatter = 73~K), luminosity ($0.049\,L_{\odot}$), and radius ($0.027\,R_{\odot}$). Our repeated measurements of these $H$-band features (Figure \ref{nferrcomp}) suggests errors of $0.1-0.2$~\AA, which corresponds to measurement errors of approximately 90~K, $0.04\,R_{\odot}$, and $0.07\,\log{L_{\odot}}$. 

To estimate the RV measurement uncertainty, we report the standard deviation of the RVs measured from each spectral order. This empirical measure of uncertainty is conveniently available for all targets, and avoids potential issues of RV variability which may occur using the epoch-to-epoch RMS differences. The median RV spread is $\sim5.7$~km~s$^{-1}$. We note that a significant component of this RV spread is contributed by systematic offsets of $\sim2-15$~km~s${-1}$ between spectral orders. We found these shifts to be variable depending on the choice of RV template spectrum, suggesting issues with the wavelength calibration, line shapes, or PSF changes. We also found overall RV differences of a similar magnitude ($\pm20$~km~s$^{-1}$) for each star depending on the template used, suggesting that our RV precision is dominated by these systematic issues. This also suggests that the $\sim7$~km~s$^{-1}$ offset between our RVs and those in the literature could be an artifact of the same systematic wavelength calibration issue.

Alternatively, it is possible to calculate measurement uncertainties based on the errors reported by the SpeXTool pipeline, which reports a combined noise that includes photon noise, read noise, and residual background noise components. Measurement uncertainties can then be estimated using Monte Carlo realizations of the noise. This process does not naturally include correlated noise effects like those arising from the telluric correction, although an ad-hoc technique is presented in N14 that mimics the effects of correlated noise. Briefly, this procedure involves convolving the noise ``spectrum" with a 1.5 pixel gaussian, which is a good approximation for the actual autocorrelation function for observed spectra. This procedure effectively spreads the noise for each pixel over the neighboring pixels, mimicking the effect of the systematic correlated errors. Applying this technique to our measurements yields error estimates that are consistently smaller than the observed epoch-to-epoch RMS differences, so we elect to report the more conservative RMS numbers as our estimated measurement uncertainties.

We note that the \ewna{} and \ewca{} are thought to be sensitive to chromospheric activity \citep{Kafka:2006kw,Barnes:2014bj,Terrien:2015eu}, and may be variable on the timescales (days-months) of our repeated observations. The epoch-to-epoch RMS differences may therefore be an overestimate of our measurement error for these features. We also estimate a lower limit of $\sim0.1$~\AA~on the measurement uncertainties for these features using a 100-trial Monte Carlo technique based on the errors reported by the SpeXTool pipeline. Following N14, we include an ad hoc correction to account for the effects of correlated noise. The true measurement errors for \na{} and \ca{} likely lie in the range bracketed by these estimates, $0.1-0.5$~\AA.

\section{Catalog Description}
We provide here a brief description of the spectra of the M dwarfs observed in this survey, as well as the set of measured and compiled values for each M dwarf. 

\subsection{The Spectra}
The spectra for each of the 886 M dwarfs are available as a \texttt{gzipped} archive of standard FITS tables hosted at the Penn State Scholarsphere data hosting service\footnote{\url{https://scholarsphere.psu.edu/files/5712mq85s\#.VWNMS9rBzGc}}. The file for each spectrum is identified by the 2MASS ID of the target star. The FITS headers contain the standard information output by the SpeXTool pipeline, including observation meta-data (instrument configuration, airmass, observing time, etc.) and the SpeXTool-generated file history. We append to this the measured RV shifts and barycentric correction for each spectrum, which can be used to velocity shift the spectra into a common reference frame.

The spectra themselves are in the standard format described fully in the SpeXTool documentation \citep{2004PASP..116..362C}\footnote{Up-to-date manuals are available at the IRTF-SpeX website: \url{http://irtfweb.ifa.hawaii.edu/~spex/}}. Briefly, each spectrum is a three-dimensional array, with the following axes:
\begin{enumerate}
	\item Order number ($3-8$).
	\item Data type (wavelength, flux, or variance). Wavelength is in vacuum, and is in the observer's reference frame.
	\item Pixel number.
\end{enumerate}

\subsection{The Catalog}
Table \ref{tablecols} describes the columns included in our catalog. There is one line per star, including astrometry and photometry compiled from the literature, measurements of spectral features, and derived stellar parameters. The full machine-readable table is available in the online edition of The Astrophysical Journal Supplement.

\begin{deluxetable*}{llc} 
	\tablewidth{0pc} 
	\tablecaption{M Dwarf Catalog Column Descriptions\tablenotemark{a} \label{tablecols}} 
	\tablehead{ 
		\colhead{Name} & \colhead{Description} & \colhead{Units} \\ 
	} \\ 
	\startdata 
	ID           & 2MASS Identifier & \ldots      \\ 
	RADEG        & Right Ascension in decimal degrees (J2000) & deg      \\ 
	DEDEG        & Declination in decimal degrees (J2000) & deg      \\ 
	JMAG         & 2MASS $J$ magnitude & mag      \\ 
	HMAG         & 2MASS $H$ magnitude & mag      \\ 
	KSMAG        & 2MASS $K_s$ magnitude & mag      \\ 
	E\_JMAG       & Error in 2MASS $J$ magnitude & mag      \\ 
	E\_HMAG       & Error in 2MASS $H$ magnitude & mag      \\ 
	E\_KSMAG      & Error in 2MASS $K_s$ magnitude & mag      \\ 
	2MQF         & 2MASS quality flag & \ldots      \\ 
	AMAG         & UCAC4 $A$ magnitude & mag      \\ 
	E\_AMAG       & Error in UCAC4 $A$ magnitude & mag      \\ 
	PMRA         & Proper motion in Right Ascension & mas/yr   \\ 
	PMDEC        & Proper motion in Declincation & mas/yr   \\ 
	R\_PM         & Reference for Proper Motion & \ldots      \\ 
	PLX          & Parallax & mas      \\ 
	E\_PLX        & Error in parallax & mas      \\ 
	R\_PLX        & Reference for parallax & \ldots      \\ 
	VMAG         & $V$ magnitude from UCAC4 & mag      \\ 
	RV           & Radial Velocity & m/s      \\ 
	E\_RV         & Error in Radial Velocity (scatter among spectral orders) & m/s      \\ 
	BC           & Barycentric Correction & m/s      \\ 
	MH2OJ        & H$_2$O-J index defined by Mann et al. 2013a & \ldots      \\ 
	CH2OH        & H$_2$O-H index defined by Covey et al. 2010 & \ldots      \\ 
	RH2OK2       & H2O-K2 index defined by Rojas-Ayala et al. 2012 & \ldots      \\ 
	MJI          & J index defined by Mann et al. 2013b & \ldots      \\ 
	MHI          & H index defined by Mann et al. 2013b & \ldots      \\ 
	MKI          & K index defined by Mann et al. 2013b & \ldots      \\ 
	NSPT         & M Spectral Subtype defined in Newton et al. 2014 & \ldots      \\ 
	NSPT\_PMSU    & PMSU Spectral Type defined in Newton et al. 2014 & \ldots      \\ 
	MTEFFJ       & Effective Temperature from J index defined in Mann et al. 2013 & K        \\ 
	MTEFFH       & Effective Temperature from H index defined in Mann et al. 2013 & K        \\ 
	MTEFFK       & Effective Temperature from K index defined in Mann et al. 2013 & K        \\ 
	MFEHJ        & [Fe/H] from $J$-band defined in Mann et al. 2013 & \ldots      \\ 
	MFEHH        & [Fe/H] from $H$-band defined in Mann et al. 2013 & \ldots      \\ 
	MFEHK        & [Fe/H] from $K$-band defined in Mann et al. 2013 & \ldots      \\ 
	MMHJ         & [M/H] from $J$-band defined in Mann et al. 2013 & \ldots      \\ 
	MMHH         & [M/H] from $H$-band defined in Mann et al. 2013 & \ldots      \\ 
	MMHK         & [M/H] from $K$-band defined in Mann et al. 2013 & \ldots      \\ 
	MFEHL        & [Fe/H] for late-type M dwarfs defined in Mann et al. 2014 & \ldots      \\ 
	NFEH         & [Fe/H] defined in Newton et al. 2014 & \ldots      \\ 
	T12FEHH      & [Fe/H] from $H$-band defined in Terrien et al. 2012 & \ldots      \\ 
	EWNA         & Equivalent Width of 820nm \ion{Na}{1} doublet & 0.1nm    \\ 
	EWCA         & Equivlanet Width of 860nm \ion{Ca}{2} triplet & 0.1nm    \\ 
	NMG148       & Equivalent Width of 1.48 $\mu$m Mg feature defined in Newton et al. 2015 & 0.1nm    \\ 
	NMG150       & Equivalent Width of 1.50 $\mu$m Mg feature defined in Newton et al. 2015 & 0.1nm    \\ 
	NK151        & Equivalent Width of 1.51 $\mu$m K feature feature defined in Newton et al. 2015 & 0.1nm    \\ 
	NMG157       & Equivalent Width of 1.57 $\mu$m Mg feature feature defined in Newton et al. 2015 & 0.1nm    \\ 
	NSI158       & Equivalent Width of 1.58 $\mu$m Si feature feature defined in Newton et al. 2015 & 0.1nm    \\ 
	NCO161       & Equivalent Width of 1.61 $\mu$m CO feature feature defined in Newton et al. 2015 & 0.1nm    \\ 
	NCO162       & Equivalent Width of 1.62 $\mu$m CO feature feature defined in Newton et al. 2015 & 0.1nm    \\ 
	NAL167A      & Equivalent Width of 1.67 $\mu$m Al-a feature feature defined in Newton et al. 2015 & 0.1nm    \\ 
	NAL167B      & Equivalent Width of 1.67 $\mu$m Al-b feature feature defined in Newton et al. 2015 & 0.1nm    \\ 
	NMG171       & Equivalent Width of 1.71 $\mu$m Mg feature feature defined in Newton et al. 2015 & 0.1nm    \\ 
	N15TEFF      & Effective Temperature from Newton et al. 2015 calibration & K        \\ 
	N15RAD       & Stellar Radius from Newton et al. 2015 calibration & solRad   \\ 
	N15LOGL      & Logarithm of stellar luminosity from Newton et al. 2015 calibration & [solLum] \\ 
	NOTES        & Notes & \ldots      
	\enddata 
	\tablenotetext{a}{Table 4 is published in its entirety in the electronic edition of The Astrophysical Journal Supplement}
\end{deluxetable*}

\section{General Discussion}
\label{gendisc}
\begin{figure}
\begin{center}
\includegraphics[scale=0.35]{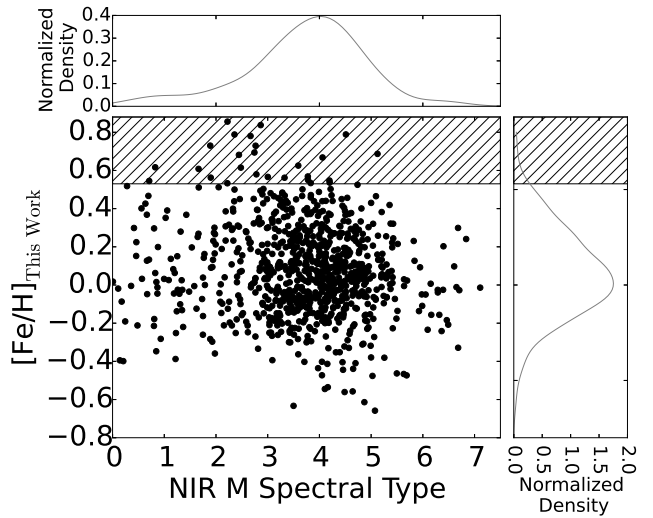}
\caption{The spectral type and [Fe/H] distributions for all M dwarfs observed in our survey. Our [Fe/H] values are the average of the $J,H,K_s$-band calibrations from M13a for targets with spectral type M5 and earlier, and the [Fe/H] derived from $K_s$ band features using the technique from M14 for targets with spectral type later than M5. The hatched region shows [Fe/H] values that exceed the calibrated range. The projection plots show a kernel density estimation for each parameter, clarifying the peaks, widths, and tails of each distribution.
\label{feh_spt}}
\end{center}
\end{figure}

The overall distribution of our sample in NIR spectral type and [Fe/H] is shown in Figure \ref{feh_spt}. The [Fe/H] distribution of our sample is centered around solar, with a median [Fe/H] of $+0.05$ dex and a scatter of $0.23$ dex. This sample is neither volume nor magnitude-limited, but is consistent with the quoted [Fe/H] measurement uncertainty ($0.07-0.11$~dex) and volume-limited measurements of the FGK stars \citep{Casagrande:2011ji} and M dwarfs \citep{2009ApJ...699..933J} in the solar neighborhood, which find a median [Fe/H] of $\sim-0.05$~dex and a scatter of $\sim0.2$~dex. The distribution of NIR spectral types, which peaks around M4 and has a long tail toward earlier spectral types, is mostly a byproduct of our observational strategy, although the mass function of the Solar neighborhood also peaks around this spectral type \citep{2006AJ....132.2360H}. 

\subsection{Internal Comparisons}
\label{intcomp}
\begin{figure*}
	\begin{center}
		\includegraphics[scale=0.35]{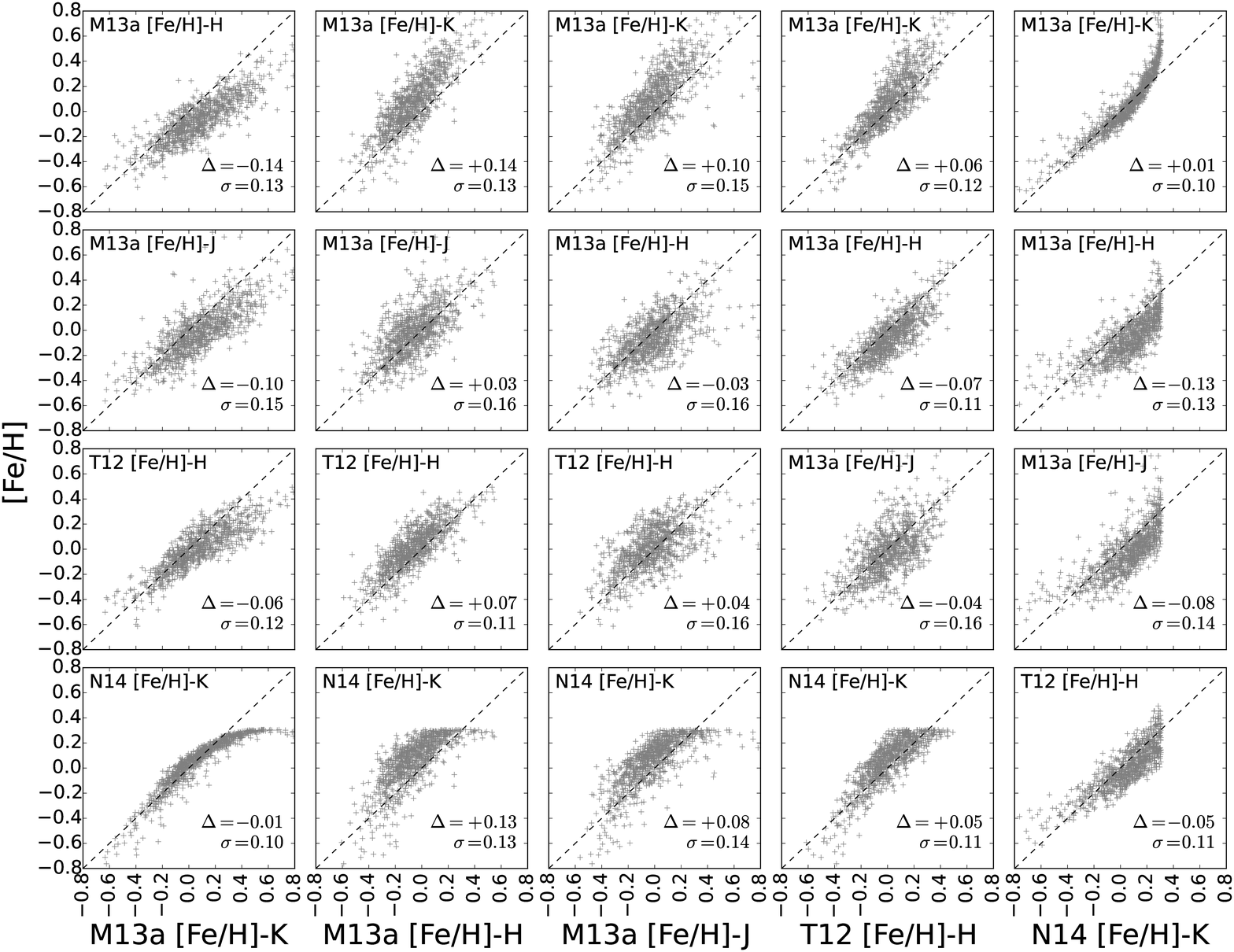}
		\caption{Comparisons between the different measures of [Fe/H] included in our catalog (vertical axis, identified within each subplot), for all M dwarfs observed. The systematic offsets between the $K_s$-band measurements and the $J,H$-band measurements are evident. Each plot lists the median offset ($\Delta$) and standard deviation ($\sigma$) of the differences, although these do not capture higher-order systematic differences.
			\label{fig_internal}}
	\end{center}
\end{figure*}

The majority of M dwarfs in our catalog have spectral types of M5 or earlier, and so can have their [Fe/H] measured with multiple techniques. We focus on the most recent of these calibrations (from M13a and N14), along with the $H$-band calibration of T12 for comparison. Our measures of [Fe/H] are compared to each other in Figure~\ref{fig_internal}, which clearly shows systematic offsets among the various calibrations of up to $0.2-0.3$~dex. In general, the $H$ and $J$-band techniques are in agreement, while the $K_s$-band techniques show good agreement up to the saturation of the N14 calibration around 0.3~dex. 

To explore whether issues with the wavelength calibration could be responsible for these systematic differences, we tested the sensitivity of the various [Fe/H] calibrations to RV shifts. We measured [Fe/H] for each target 50 times with a normal distribution of RVs centered around the original measured RV and with a spread equivalent to our estimated systematic and random measurement errors (typically 10-20 km~s$^{-1}$). Among the M13a calibrations, we found that the mean scatter in the $K_s$-band calibration was best (at $\sim0.02$~dex), compared to the $H$-band ($\sim0.03$~dex) and $J$-band ($\sim0.04$~dex). This suggests that the features chosen for the M13a $J$ and $H$-band calibrations are marginally more sensitive than the $K_s$-band features to small wavelength shifts, and the magnitude of the expected RV errors is not sufficient to explain the observed offsets. We found that an RV shift of $>100$~km~s$^{-1}$ can remove the offset between the $K_s$ and $H$-band measurements, although this causes the $J$-band measurements to diverge further. In any case, such a large shift is not justified given the agreement among our per-order RV measurements, those for different RV templates, and between our overall RV measurements and literature values. 

The differences among our implementations of the M13a techniques may be due to undiagnosed issues with the wavelength calibration or other unidentified differences between our reduction procedure and those of M13a (e.g.\ the telluric correction). We present measurements from each band in our catalog, and justify below the choice of the $K_s$-band calibration of M13a as our preferred abundance calibration for M1-M5 dwarfs (Section \ref{prefmeas}). 

\subsection{Comparison to Other Catalogs}
\label{extcomp}
To check our results and to probe the relationships between different measures of metallicity, we compared our values (using the calibrations of M13a, M14, and N14) to those presented by a number of recent works \citep[][ N14]{RojasAyala:2012fb,Gaidos:2014if} for common targets. \citet{RojasAyala:2012fb} rely on the strengths of the $K_s$-band \ion{Na}{1} and \ion{Ca}{1} features along with the H$_2$O-K2 index to account of \teff{}; N14 rely on the strength of the $K_s$-band \ion{Na}{1} feature alone; \citet{Gaidos:2014if} used the $V$-band spectroscopic calibration presented by M13a. 

\begin{figure}
	\begin{center}
		\includegraphics[scale=.5]{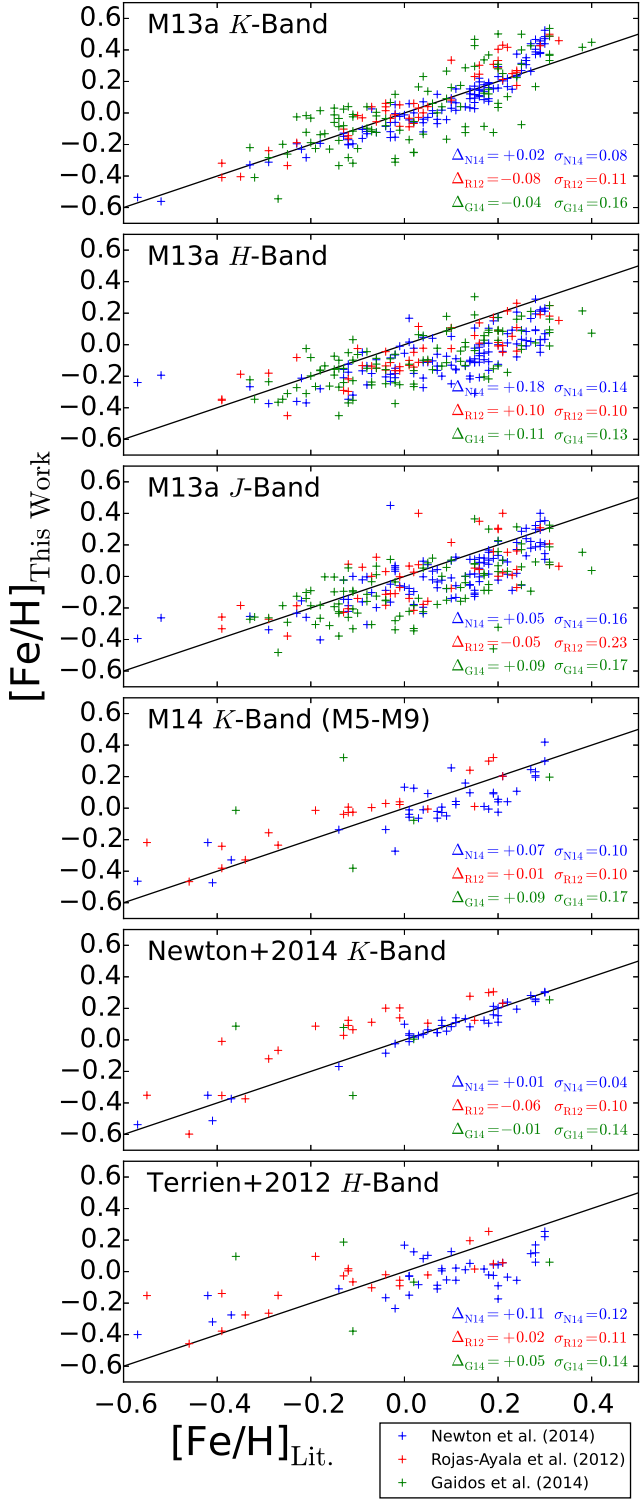}
		\caption{For stars in common with other recent surveys \citep[][N14]{RojasAyala:2012fb,Gaidos:2014if}, the comparison to [Fe/H] values we measure. Each plot lists the median offset ($\Delta$) and standard deviation ($\sigma$) of the differences, although these do not capture higher-order systematic differences.
			\label{fehcomp}}
	\end{center}
\end{figure}

The comparison of our values to those from other works is shown in Figure \ref{fehcomp}, along with the median and scatter in each case. Our application of the M13a $K_s$-band [Fe/H] calibration shows good agreement with these other works, although we find systematically higher [Fe/H] for those stars with the highest [Fe/H] values compared to values reported in \citet{RojasAyala:2012fb} and (N14). This behavior may be a byproduct of the features used in each case, whose sensitivity to [Fe/H] can saturate (as noted in N14). Particularly, the Na and Ca $K_s$-band features used in \citet{RojasAyala:2012fb}, N14, and M14 are known to saturate as [Fe/H] approaches 0.5 dex. 

The $H$ and $J$-band calibrations are known to have larger scatter (M13a), and this is borne out in our comparisons of these measurements to other works. In both cases, the [Fe/H] we measure for stars with Solar [Fe/H] and higher is systematically $0.1-0.2$~dex lower than that reported in other works. This offset, which is not present for the $K_s$-band [Fe/H] values, may result from structure remaining in the residuals of these calibrations as shown in e.g.\ Figure~5 of M13a.

There are several M dwarfs for which we measure later spectal type than M5, which also have [Fe/H] reported in other works. For these stars, the N14 and \citet{RojasAyala:2012fb} calibrations appear to perform well compared to the M14 calibration, likely due to their common reliance on the $K_s$-band \ion{Na}{1} feature. 

Our application of the N14 [Fe/H] calibration is consistent with the published [Fe/H] measurements in their work, verifying our implementation. For targets in common with \citet{RojasAyala:2012fb}, our [Fe/H] measurements are systematically higher, consistent with the offset for later spectral types observed in Figure~18 of N14.

\begin{figure}
	\begin{center}
		\includegraphics[scale=.4]{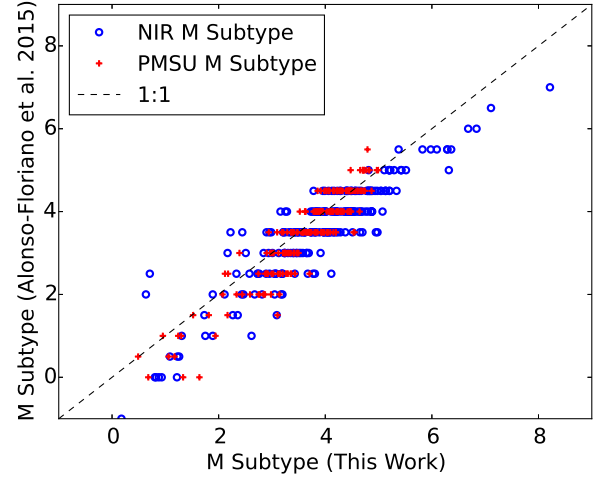}
		\caption{For stars in common the CARMENES target catalog \citep{FJAlonsoFloriano:2015bd}, the respective measures spectral type.
			\label{ccomp}}
	\end{center}
\end{figure}

A similar catalog to the one presented here is that being developed for the CARMENES M dwarf exoplanet survey \citep{Quirrenbach:2014fj}, for which an extensive optical spectroscopic survey has been carried out \citep{FJAlonsoFloriano:2015bd}. There are 276 targets in the CARMENES catalog with coordinate matches in our catalog to within 5''. We plot the respective measures of composition and spectral type in Figure \ref{ccomp}. As an indication of abundance, \citet{FJAlonsoFloriano:2015bd} measure the [Fe/H]-sensitive $\zeta$ parameter \citep[based on TiO and CaH absorption in the optical,][]{2007ApJ...669.1235L,2012AJ....143...67D}. We find that $\zeta$ is weakly correlated with our measured [Fe/H] (Kendall's $\tau=0.25$, two-sided $p=1.3\times 10^{-8}$). This is unsurprising since $\zeta$ is a coarse indicator of abundance, which has been primarily used for discriminating between dwarfs and subdwarfs. For all targets in the overlap sample, our [Fe/H] measurements and the $\zeta$ measurements of \citet{FJAlonsoFloriano:2015bd} are both consistent with dwarf compositions. To measure spectral type, \citet{FJAlonsoFloriano:2015bd} use a suite of spectral standard template matching and spectral index based techniques, and report spectral types to 0.5 subtypes. Our spectral type estimates broadly agree with those of \citet{FJAlonsoFloriano:2015bd}: our NIR M subtype measurements have a median offset of 0.37 with a residual standard deviation of 0.52, while our PMSU spectral types have a median offset of 0.12 subtypes and a residual standard deviation of 0.41. \citet{FJAlonsoFloriano:2015bd} use PMSU standards for much of their reference set, so it is expected that the PMSU measurements should show better agreement.

\subsection{Preferred Measure of [Fe/H] for Early-Mid M Dwarfs}
\label{prefmeas}
In light of the systematic offsets shown in Sections \ref{intcomp} and \ref{extcomp}, it is useful to choose and justify a preferred measure of [Fe/H] for M1-M5 dwarfs. We choose the $K_s$-band calibration of M13a for the following reasons:
\begin{itemize}
	\item The breadth and number of calibrators exceeds those of other similar calibrations \citep[T12,][]{RojasAyala:2012fb,2014AJ....147...20N}.
	\item The better agreement between the $K_s$-band measurements and other literature measurements of the star (Section \ref{extcomp}), which are themselves based on $K_s$-band features \citep[][, N14]{RojasAyala:2012fb} and visible features \citep{Gaidos:2014if}.
	\item The stability against small RV shifts of the $K_s$-band measurements compared to those of the $J$ and $H$-band of M13a.
\end{itemize}  

\subsection{[Fe/H] and Stellar Kinematics}
The stellar population of the Galaxy comprises multiple sub-populations with distinct kinematic, spatial, and compositional characteristics. These populations are often grouped into three components---the thin disk, the thick disk, and the Galactic halo \citep[although the disk populations may be better-represented by a continuous distribution of populations than a bi-modal distribution,][] {Bovy:2012du}. Large scale surveys of FGK stars and luminous evolved stars have found rich information about the history and evolution of the Galaxy within the detailed abundances of these different populations \citep[e.g.][]{Bensby:2003fk,Bensby:2014gi}. Similar large-scale studies of M dwarfs \citep[e.g.][]{2011AJ....141...97W,2011AJ....141...98B} have also been carried out, and measurements of metallicity-sensitive band indices have enabled the identification and study of the oldest metal-poor M subdwarf populations \citep[e.g.][]{Bochanski:2007ij}. Refined metallicity measurement techniques are under development, and have demonstrated great potential for application with these large M dwarf catalogs \citep{2010AJ....139.2566D,2012AJ....143...67D}. Precise metallicity measurements of the M dwarfs in the solar neighborhood reveal correlations between composition and stellar kinematic and activity characteristics that are consistent with the increasing age and decreasing level of activity from the thin disk to the thick disk \citep{RojasAyala:2012fb}. These results suggest the promise of precise abundance measurements in these long-lived stars.

\begin{figure*}
\begin{center}
\includegraphics[scale=0.35]{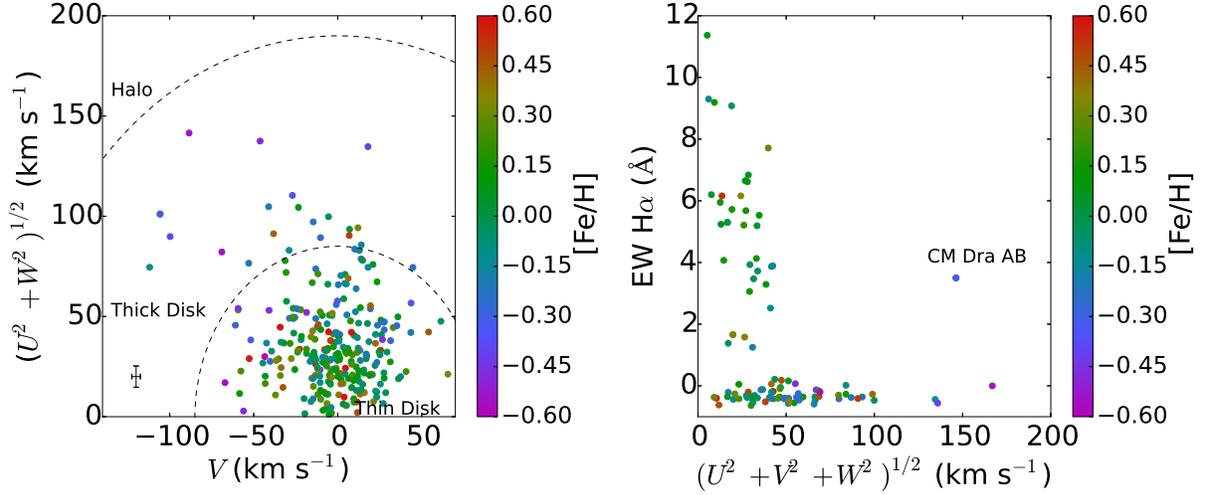}
\caption{\textit{(Left):} The Toomre diagram for survey targets with measured parallax. The halo, thick disk, and thin disk regions are denoted. There are no metal-rich M dwarfs with thick-disk kinematics in our sample, consistent with expectations. A typical error bar is plotted in the lower left. \textit{(Right):} The total Galactocentric velocity versus H$\alpha$ EW (positive indicates emission and therefore more choromospheric activity). Indicators of both [Fe/H] and activity (H$\alpha$) suggest that our catalog probes both the thick and thin disks.
\label{uvw}}
\end{center}
\end{figure*}

We explored whether the [Fe/H] values in were consistent with the expected trends, as a check on our [Fe/H] values and to probe for outliers and systematics. We calculated Galactocentric UVW motions for all of our targets with parallax measurements. Combining these parallaxes with proper motion values from LG11 and LSPM-N and RV measurements from the literature where available (or our RV measurements where literature RVs were unavailable), we have 283 targets with relatively precise UVW velocities (parallax error $<10\%$, literature RV or measured RV error $<10$~km~s$^{-1}$, no indications of binarity). These velocities are shown in the Toomre diagram in the left panel of Figure \ref{uvw}, along with rough demarcations of the thin disk, thick disk, and halo \citep{Fuhrmann:2004js}. The Toomre diagram is effectively a projection of the rotational kinetic energy of each star relative to the local standard of rest, with stars of higher energies at increasing radii \citep{Sandage:1987el}. Consistent with expectations from galactic kinematics, those targets with the highest [Fe/H] values are kinematically within the thin disk region, and targets with lower [Fe/H] values are spread across the entire range of velocities. The right panel of Figure \ref{uvw} shows the total Galactocentric velocity plotted against H$\alpha$ EW, a common activity indicator for M dwarfs \citep[e.g.][]{Reid:1995kw,West:2004kd,2009AJ....137.3297W}. The concentration of chromospherically active M dwarfs at low Galactocentric velocities, compared with the absence of active M dwarfs at higher velocities, suggests that we are indeed probing both the thick and thin disks, consistent with the chemical and kinematic properties of the M dwarfs in our catalog. The boundary between the thick and thin disk populations is not discrete \citep{Ivezic:2008fd}, and there are indications that the compositional gradient between the thick and thin disks should be less than 1 dex \citep{Bochanski:2007ij}, consistent with Figure \ref{uvw}. 

The most significant outlier in these plots is CM Dra AB, a well-studied short-period eclipsing binary M dwarf system \citep[e.g.][]{2009ApJ...691.1400M,Terrien:2012hj,Feiden:2014dn}. Constraints from the cooling age of a nearby co-moving white dwarf \citep{2009ApJ...691.1400M,Feiden:2014dn} support the old age of this system, consistent with its high velocity, low [Fe/H], and possible $\alpha$-enrichment \citep{Feiden:2014dn,Terrien:2015eu}. That this system is chromospherically active at its old age is not surprising due to its short period ($1.27$~d) and the tidal interactions that have presumably taken place.

\subsection{Metallicity and Stellar Colors}
The infrared colors of M dwarfs have long been known to show sensitivity to stellar metallicity. This is in large part due to two primary sources of continuum opacity: H$^-$ absorption and H$_2$O absorption. Lower-metallicity stars have higher H$^-$ and H$_2$O absorption, leading to a loss of flux especially in the $H$-band \citep{1976A&A....48..443M,1992ApJS...82..351L}. The effects of metallicity on infrared colors have been exploited by many authors to construct photometric metallicity calibrations \citep[][M13a, N14]{2012AJ....143..111J}, which are complementary (observationally cheaper but less precise) to the spectroscopic [Fe/H] provided in this work. 

\begin{figure}
	\begin{center}
		\includegraphics[scale=0.35]{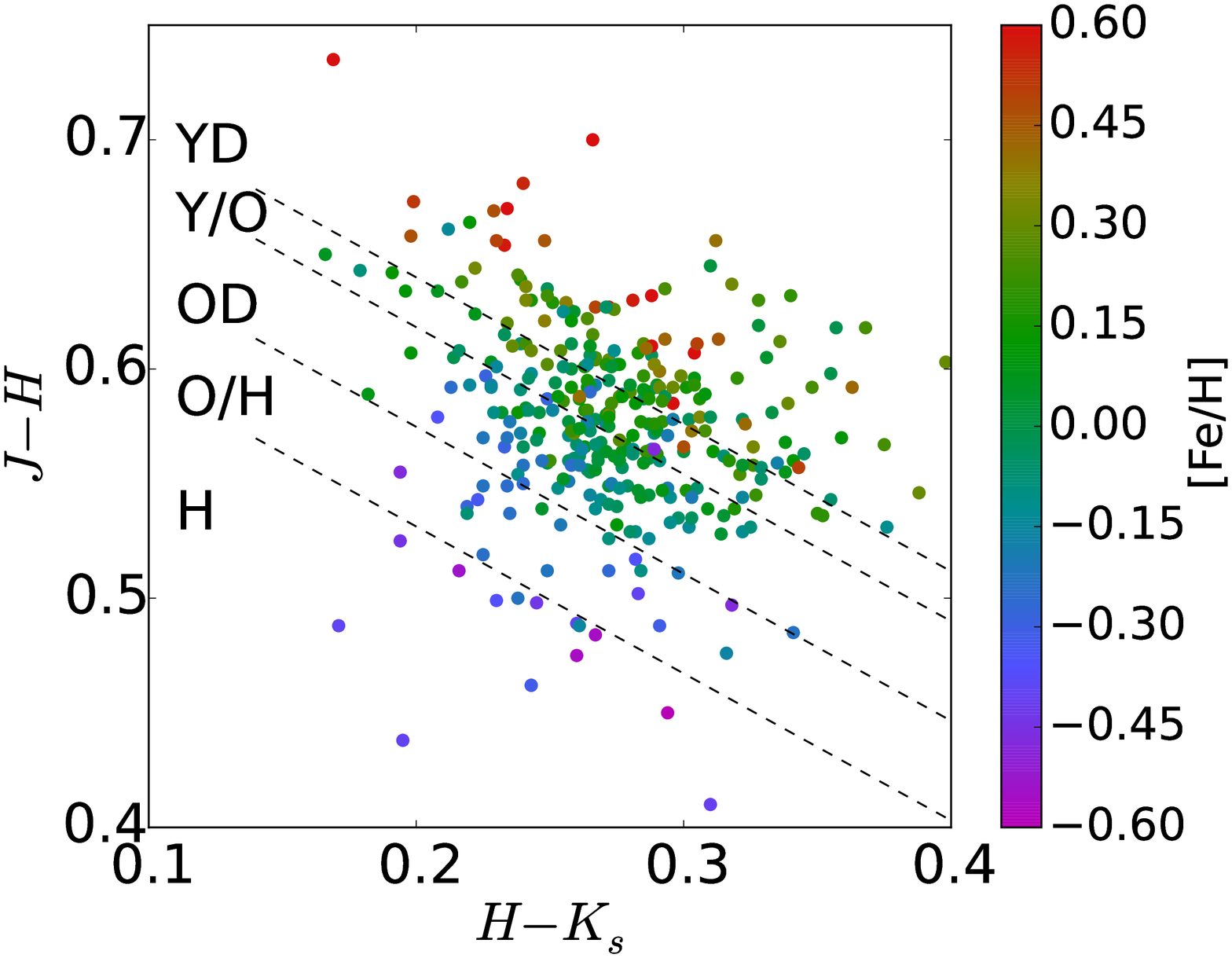}
		\caption{A NIR color-color diagram for survey targets with high-quality (AAA) 2MASS photometry and no indications of multiplicity. The position in $(J-H, H-K_{s})$ space is well-correlated with metallicity, as noted in \citep{1992ApJS...82..351L}.
			\label{colfeh}}
	\end{center}
\end{figure}

As another check of the integrity of our catalog of metallicities, we show in Figure \ref{colfeh} the $(J-H,H-K_s)$ colors along with the [Fe/H] values for each of our targets. We confirmed that our data show a tight relation between position in the $(J-H, H-K_{s})$-space and [Fe/H], a connection established through examination of kinematically-defined populations in \citet{1992ApJS...82..351L}. Also shown are the kinematic age boundaries derived in \citet{1992ApJS...82..351L}, transformed to the 2MASS photometric system using the updated\footnote{\url{http://www.astro.caltech.edu/~jmc/2mass/v3/transformations/}} 2MASS color transformations \citep{Carpenter:2001fv}.

\subsection{Planet Hosts}
Stellar composition is of fundamental importance in studies of exoplanets. The properties of observational relationships relating to composition, including the giant planet-metallicity \citep{2005ApJ...622.1102F} and the planet mass-metallicity \citep{Ghezzi:2010hu} correlations, provide unique insight into the environments and mechanisms of planet formation. Moreover, reliable planet host star compositions are necessary in order to apply empirical relationships to derive the stellar luminosities and temperatures \citep{Boyajian:2012eu,Muirhead:2014gw,2015ApJ...804...64M}, and hence the location of the Habitable Zone \citep[e.g.][M13b]{Muirhead:2012dxa}. 

This breadth of impact motivates the empirical and theoretical efforts to define an M dwarf metallicity scale. Multiple groups have already shown that the giant planet-metallicity correlation extends to the M dwarf regime \citep{2009ApJ...699..933J,2013A&A...551A..36N}. A well-calibrated and precise metallicity scale is crucial for these and future studies of M dwarf planet populations. The metallicity values in this catalog are derived from the most precise and well-calibrated scales available for M dwarfs, and it is valuable to compare our results with those of previous studies for planet-hosting M dwarfs.

Our catalog includes 16 M dwarfs that are known to be planet hosts, and in Table \ref{ptable} we show our measured [Fe/H], [M/H], and H$_2$O-K2 based spectral types. Generally, we measure metallicities consistent with those of previous NIR spectroscopic studies. In many cases our [Fe/H] measurements are substantially different from previous photometric estimates, which is expected as those calibrations are inherently less precise and use fewer calibrators than the relations we employed. 

\begin{deluxetable*}{lccccccccccccc}
\tablewidth{0pc}
\tablecaption{Measured parameters for known planet hosts \label{ptable}}
\tablehead{
\colhead{Name} & \colhead{B05\tablenotemark{a}} & \colhead{JA09\tablenotemark{a}} & \colhead{W09\tablenotemark{a}} &  \colhead{SL10\tablenotemark{a}} & \colhead{RA10\tablenotemark{a}} & \colhead{RA12\tablenotemark{a}} &  \colhead{N12\tablenotemark{a}} & \colhead{T12\tablenotemark{a}} & \colhead{N14\tablenotemark{a}} & \colhead{other} &  \multicolumn{3}{c}{This Work}   \\
 & \colhead{[Fe/H]} & \colhead{[Fe/H]} & \colhead{[Fe/H]} & \colhead{[Fe/H]} & \colhead{[Fe/H]} & \colhead{[Fe/H]} & \colhead{[Fe/H]} & \colhead{[Fe/H]} & \colhead{[Fe/H]} & & \colhead{[Fe/H]\tablenotemark{b}} & \colhead{[M/H]\tablenotemark{c}} & \colhead{NIR SpT} } 

\startdata      
GJ 1214  &  \ldots  & $+0.03$ & \ldots  & $+0.28$ & $+0.39$ & $+0.20$ & $+0.03$ & \ldots  & $+0.05$ &  \ldots & $+0.40$  &  $+0.27$  &   M4.5 \\ 
HIP 57050  & $-0.02$ & $+0.32$ & $+0.07$ & \ldots  & $+0.12$ & $+0.04$ & \ldots  & $+0.05$ & \ldots  &  \ldots & $+0.04$  &  $-0.03$  &   M4.0 \\    
GJ 876  & $+0.03$ & $+0.37$ & $+0.02$ & $+0.23$ & $+0.43$ & $+0.19$ & $+0.14$ & $+0.11$ & $+0.14$ & \ldots &  $+0.31$  &  $+0.19$  &   M3.5 \\    
GJ 317  &  \ldots  & \ldots  & \ldots  & \ldots  & \ldots   & \ldots & $+0.22$ & $+0.31$ & $+0.22$ &  \ldots & $+0.43$  &  $+0.30$  & M3.5  \\	
GJ 581 & $-0.25$ & $-0.10$ & $-0.10$ & $-0.22$ & $-0.02$ & $-0.10$ & $-0.18$ & $-0.09$ & $-0.20$ &  \ldots & $-0.02$  &  $-0.03$  &   M3.0 \\  
GJ 179  & \ldots  & $+0.30$ & $+0.02$ & $+0.20$ & \ldots  & $+0.23$ & $+0.14$ & $+0.18$ & $+0.12$ &  \ldots & $+0.25$  &  $+0.12$  &   M3.0 \\      
HIP 79431  & $+0.16$ & $+0.52$ & \ldots  & $+0.35$ & $+0.60$ & $+0.46$ & \ldots  & $+0.50$ & \ldots  &   \ldots & $+0.78$\tablenotemark{d}  &  $+0.55$  &   M2.5 \\     
GJ 436  & $-0.03$ & $+0.25$ & $-0.05$ & $+0.10$ & $+0.00$ & $+0.05$ & $+0.01$ & $+0.02$ & $-0.03$ &  \ldots & $+0.00$  &  $-0.03$  &   M2.5 \\      
GJ 849  & $+0.14$ & $+0.58$ & $+0.20$ & $+0.41$ & $+0.49$ & $+0.31$ & $+0.24$ & $+0.22$ & $+0.22$ &  \ldots & $+0.50$  &  $+0.33$  &   M2.5 \\        
GJ 3470  &  \ldots  & \ldots  & \ldots  & \ldots  & \ldots  & \ldots  & \ldots  & \ldots & \ldots  &  $+0.20$\tablenotemark{e} & $+0.27$  &  $+0.15$  &   M2.0  \\         
GJ 15A  &  \ldots  & \ldots  & \ldots  & \ldots  & \ldots  & \ldots  & \ldots  & \ldots & \ldots  &  $-0.36$\tablenotemark{f} & $-0.28$  &  $-0.17$  &   M2.0  \\             
GJ 649  & $-0.18$ & $+0.04$ & $-0.13$ & $-0.03$ & $+0.14$ & $-0.04$ & \ldots  & $-0.05$ & \ldots  &  \ldots & $+0.04$  &  $-0.01$  &   M1.5 \\      
GJ 433  &  \ldots  & \ldots  & \ldots  & \ldots  & \ldots  & \ldots & $-0.13$ & \ldots & $-0.17$ &  \ldots & $-0.03$  &   $-0.06$  &   M1.5  \\         
Kepler-138  &  \ldots  & \ldots  & \ldots  & \ldots  & \ldots  & \ldots  & \ldots  & \ldots  & \ldots  &  $-0.18$\tablenotemark{g} & $-0.21$  &  $-0.16$  &   M1.0  \\       
WASP-80  &  \ldots  & \ldots  & \ldots  & \ldots  & \ldots  &  \ldots  & \ldots  & \ldots  & \ldots  &  $-0.14$\tablenotemark{h} & $+0.13$  &  $+0.07$  &   M0.5 \\  
WASP-43  &  \ldots  & \ldots  & \ldots  & \ldots  & \ldots  &  \ldots  & \ldots  & \ldots  & \ldots  &  $+0.01$\tablenotemark{i} & $+0.40$  &  $+0.28$  &   M0.5 \enddata   
\tablenotetext{a}{B05:\citet{2005A&A...442..635B}, JA09:\citet{2009ApJ...699..933J}, W09:\citet{2009PASP..121..117W}, SL10:\citet{2010A&A...519A.105S}, RA10:\citet{RojasAyala:2010ht}, RA12:\citet{RojasAyala:2012fb}, N12:\citet{2012A&A...538A..25N}, T12:\citet{2012ApJ...747L..38T}, N14:\citet{Neves:2014jj}}
\tablenotetext{b}{Uncertainties in [Fe/H] are approximately 0.11 dex}
\tablenotetext{c}{Uncertainties in [M/H] are approximately 0.10 dex}
\tablenotetext{d}{This star is well outside the calibrated range ($-1.04 < \mathrm{[Fe/H]} < +0.56$) of [Fe/H].}
\tablenotetext{e}{\citep{2013ApJ...768..154D}}
\tablenotetext{f}{\citep{Neves:2013cm}}
\tablenotetext{g}{[M/H] value from \citet{Muirhead:2012dxa}}
\tablenotetext{h}{\citep{Triaud:2013ek}}
\tablenotetext{i}{\citep{Chen:2014hu}}
\end{deluxetable*}

\section{Conclusion}
\label{conclusion}
We have presented NIR spectra and spectroscopic parameters for 886 nearby M dwarfs. This catalog represents the largest such compilation of M dwarf abundances, \teff{}-sensitive indices, and NIR spectra yet published, and includes:
\begin{itemize}
	\item {[}Fe/H] and [M/H] measurements (uncertainty $\sim0.1$~dex) for M1-M5 dwarfs using $J$, $H$, and $K_s$-band features, derived from multiple published calibrations.
	\item {[}Fe/H] with similar uncertainties derived from $K_s$-band features for M5-M8 dwarfs.
	\item Measurements of $J,H,K_s$-band H$_2$O-based and other indices which are known to track \teff{}.
	\item Measurements of strengths of the 820~nm \na{} doublet and the 860nm \ca{} triplet, which are sensitive to activity and surface gravity.
	\item Stellar photometry and astrometry from the literature.
	\item RVs (uncertainty $\sim10$~km~s$^{-1}$).
\end{itemize}

We performed cross-checks among the different [Fe/H] calibrations within our catalog and in the literature. We uncovered some systematic issues with the implementations of the $J$ and $H$-band calibrations, which led us to select the $K_s$-band calibration of M13a as the preferred measure of [Fe/H] for M1-M5 dwarfs where multiple calibrations exist. Using this [Fe/H], we confirmed the expected trends with Galactocentric velocity activity levels, and stellar colors. We also provide updated abundances measurements for many of the known M dwarf exoplanet hosts.

The original motivation for this survey was to support upcoming planet surveys targeting M dwarfs, including ground-based RV surveys \citep{Mahadevan:2014be,Quirrenbach:2014fj,Thibault:2012fb,Kotani:2014fe} and space-based transit surveys \citep{Rauer:2014kx,Ricker:2015ie}, to enable both survey target planning and rapid interpretation of results (e.g.\ whether a planet is in the Habitable Zone of its host star). This catalog will also be useful for other studies of the nearby M dwarfs, and has already proven its value in this context. Data from this catalog enabled insight into the tension between observations and model predictions for the eclipsing binary CM Dra \citep{Terrien:2012hj}, supported the first confirmation of the low-mass members of the Coma Berenices cluster \citep{Terrien:2014jq}, and were used to demonstrate the use of new gravity and activity-sensitive spectral indices \citep{Terrien:2015eu}. This collection of M dwarf spectra provide a rich dataset on which to build even better calibrations, or to discover new useful features. 

It is our hope that this dataset, complementing other similar recent works \citep[e.g.][]{RojasAyala:2012fb,Lepine:2013hc,2014AJ....147...20N,Gaidos:2014if,FJAlonsoFloriano:2015bd}, will facilitate the work of other groups working to understand the nearby M dwarfs, and to further our understanding of these most common stars and their planets.

\bigskip

We thank Dr. Kevin Luhman for obtaining observations of a subset of these targets. We also thank the referee, Dr. Andrew Mann, for a thorough and insightful review.

This work was partially supported by funding from the Center for Exoplanets and Habitable Worlds. The Center for Exoplanets and Habitable Worlds is supported by the Pennsylvania State University, the Eberly College of Science, and the Pennsylvania Space Grant Consortium. This work was also partially supported by the Penn State Astrobiology Research Center and the National Aeronautics and Space Administration (NASA) Astrobiology Institute. We acknowledge support from NSF grants AST 1006676, AST 1126413, and AST 1310885 in our pursuit of precision radial velocities in the NIR. This research has made use of the SIMBAD database, operated at CDS, Strasbourg, France. This publication makes use of data products from the Two Micron All Sky Survey, which is a joint project of the University of Massachusetts and the Infrared Processing and Analysis Center/California Institute of Technology, funded by the National Aeronautics and Space Administration and the National Science Foundation. 

The authors wish to recognize and acknowledge the very significant cultural role and reverence that the summit of Mauna Kea has always had within the indigenous Hawaiian community.  We are most fortunate to have the opportunity to conduct observations from this mountain.

\bibliographystyle{apj}
%\bibliography{refs}

\end{document}